\documentclass[12pt,fleqn]{article}

\usepackage[cp1251]{inputenc}
\begin{document}
\large

\begin{center}
{\Large \bf Existence of solutions to non-homogeneous higher order
differential equation in the Schwartz space}
\end{center}

\vskip5mm

\begin{center}
\noindent Valerii H. Samoilenko{\footnote {corresponding author}},
Yuliia I. Samoilenko \\
Department of Mathematical Physics, \\
Taras Shevchenko National
University of Kyiv, \\ vul. Volodymyrs'ka, 64, \, Kyiv, \, Ukraine, \, 01601 \\
valsamyul@gmail.com, vsam@univ.kiev.ua, yusam@univ.kiev.ua \\
Samoilenko V.~H. ORCID ID: 0000-0002-3267-2923 \\
Samoilenko Yu.I. \, ORCID ID: 0000-0002-9923-0986
\end{center}

\vskip5mm

\begin{minipage}{150mm}
{\bf Abstract.} There is studied problem on existence of solutions
to non-homogeneous differential equation of higher even order.
Similar problem arises while studying soliton and soliton-like
solutions to partial differential equations of integrable type.

By means of Fourier transform and theory of pseudodifferential
operators there is proved the theorem on necessary and sufficient
conditions on existence of solutions to linear non-homogeneous
differential equation of higher even order in the Schwartz space.
\end{minipage}

\vskip10mm


\vskip10mm


\vskip10mm

\section{Introduction and statement of problem} One of the most
important problems of the qualitative theory of differential
equations is the problem of existence and uniqueness of solutions to
the equations. Similar problems need to be studied frequently when
considering the qualitative properties of differential equations, as
well as of mathematical models of natural sciences. In this
connection it should be noted the problem on existence of particular
solutions with special properties that leads to searching the ones
in certain functional spaces \cite{GHHP}. For example, while
considering soliton solutions in hydrodynamical models there is
appeared a problem on determining solutions in space of quickly
decreasing functions. The same problems are also arisen when
constructing asymptotic soliton like solutions for different partial
differential equations of integrable type with singular
perturbations \cite{Maslov}.

While studying asymptotic solutions of soliton type to singular
perturbed Korteweg-de Vries equation with variable coefficients
\cite{Sam1} the problem on existence of a solution to the following
equation
\begin{equation} \label{eq_Shrodinger}
\frac{d^2 v}{dx^2} + q v = f, \quad x \in{\mathbf R},
\end{equation}
in Schwartz space has been generated.


On the other hand, when searching soliton solutions to the higher
Korteweg-de Vries equations and KdV-like equations there is appeared
the same problem for similar equations with differential operator of
higher order. So, there is come up the problem on finding necessary
and sufficient conditions on existence of a solution to the
following equation
\begin{equation} \label{high_eq}
L v = f, \quad x \in{\mathbf R},
\end{equation}
in the Schwartz space, where the differential operator $L$ is
written as
\begin{equation} \label{operator_L}
L = - \sum\limits_{m=1}^n a_{2m} \frac{d^{2m}}{d x^{2m}} + q
\end{equation}
with constant coefficients $ a_{2m}, m = \overline{1,n} $, and $ q $
being a function of $x$.

\section{Main result} Let $ \mathrm{S} ({\mathbf R}) $ be the Schwartz space.

The main result of the paper is the statement.

{\bf Theorem 1.} {\it Let the following conditions be fulfilled:

$1^0.$ the coefficients $ a_{2m}, m = \overline{1,n} $, are
nonnegative constants and $ a_{2n} > 0 $;

$2^0.$ $ q(x) = q_0 + q_1(x) $, where constant $q_0 <0$ and the
function $ q_1(x) \in \mathrm{S}({\mathbf R}) $;

$3^0.$ the function $ f \in \mathrm{S}({\mathbf R}) $.

If the kernel of the operator $ L : \mathrm{S}({\mathbf R}) \to
\mathrm{S}({\mathbf R}) $ is trivial, then equation (\ref{high_eq})
has a solution in the space $ \mathrm{S}({\mathbf R}) $ for any
function $ f $. 

Otherwise, if the kernel of the operator $ L: \mathrm{S}({\mathbf
R}) \to \mathrm{S}({\mathbf R}) $ is not trivial, then equation
(\ref{high_eq}) has a solution in the space $ \mathrm{S}({\mathbf
R}) $ if and only if the function $ f $ 
satisfies the condition of orthogonality in the form
\begin{equation}\label{ort_cond}
\int\limits_{-\infty}^{+\infty} f (x) v_0(x) d x = 0
\end{equation}
for any $ v_0 \in ker~L $. }

\section{Necessary definitions and statements} To prove the theorem
we need to remind some notations, definitions and results. For any
function $ h \in \mathrm{S} ({\mathbf R}) $ there is denoted the
Fourier transform as
$$
F[h](\xi) = \int\limits_{-\infty}^{+\infty} e^{- i \xi x} h(x) d x.
$$

Due to properties of the Fourier transform for any differential
operator
$$
p \left( x, \frac{d}{dx} \right) = \sum\limits_{k = 0}^n a_k(x)
\frac{d^k}{dx^k} , \quad x \in {\mathbf R},
$$
it's possible to define its action on function $ h \in \mathrm{S}
({\mathbf R}) $ as
\begin{equation} \label{symbol}
p \left( x, \frac{d}{dx} \right) h (x) = \frac{1}{2\pi}
\int\limits_{-\infty}^{+\infty} e^{i x \xi} \, p(x, \xi) F[h](\xi)
\,   d \xi .
\end{equation}

Here
$$ p(x, \xi) = \sum\limits_{k = 0}^n a_k(x) (- i\xi)^k , \quad x,
\xi \in {\mathbf R},
$$
is called a symbol of the differential operator $ p \left( x,
\frac{d}{dx} \right) $.

Let $ S^{\, m} $ be a set of symbols $ p(x, \xi) $ $ \in $ $
\mathrm{C}^{\infty} ({\mathbf R}^2) $ such that for any $ k $, $ l
\in {\mathbf N}\cup\{0\} $ the inequality
$$
\left| \, p_{(l)}^{(k)} (x, \xi) \right| \le C_{k l} \left( 1 +
|\xi|\right) ^{m - k}, \quad (x, \xi) \in {\mathbf R}^2,
$$
is true, where
$$
p_{(l)}^{(k)} (x, \xi) = \frac{\partial^{k + l}}{\partial\xi^k
\partial x^l} \, p(x, \xi), \quad (x, \xi) \in {\mathbf R}^2,
$$
and values $ C_{kl} $, $ k $, $ l \in {\mathbf N}\cup\{0\} $, are
some constants \cite{Hor}.

By $ S^{\, m}_0 $ denote a set of symbols $ p(x, \xi) \in S^m $ such
that
$$
|p(x, \xi) | \le M(x) \left(1 + |\xi| \right)^m,
$$
where value $ M(x) \to 0 $ as $ |x| \to + \infty $.

Let $ \mathrm{H}_s({\mathbf R}) $, $ s \in {\mathbf R} $, be a
Sobolev space \cite{GR1}, i.e. a space of generalized functions $ g
\in \mathrm{S}^* ({\mathbf R}) $ such that their Fourier transform $
F[g](\xi) $ satisfies condition
\begin{equation}\label{norma}
|| g ||_s^2 =  \int\limits_{-\infty}^{+\infty} (1 + |\xi|^2)^s \, \,
| F[g](\xi) |^2 \, d \xi  < \infty .
\end{equation}

It is worthy to recall the following theorem.

{\bf Theorem 2} {(Grushin, \cite{GR1}).} {\it Let $ p(x, \xi) \in
S^m $ be a symbol such that $ \partial^l p(x, \xi) /
\partial x^l \in S^{\,m}_0 $, $ l \in {\mathbf N} $, and inequality
$$
\lim\limits_{(x\overline{, \xi) \to} \infty} \frac{|p(x, \xi)|}{(1 +
|\xi|)^m}
> 0
$$
is true.

Then $ p \left( x, \frac{d}{dx}\right) : H_{s+m}({\mathbf R}) \to
H_s ({\mathbf R})$, defined through formula (\ref{symbol}), is the
Noether operator for any $ s \in {\mathbf R} $. }

\section{Proof of the main result} Proving the theorem 1 contains two
steps. Firstly, we show that the operator $ L: \mathrm{H}_{s+2n}
({\mathbf R}) \rightarrow \mathrm{H}_s ({\mathbf R})$ of form
(\ref{operator_L}) is the Noether operator for any $ s \in {\mathbf
R} $. Later we prove that the solution to equation (\ref{high_eq})
belongs to the Schwartz space $ \mathrm{S}({\mathbf R}) $.

Let us consider symbol of the differential operator $ L $ having a
form
\begin{equation} \label{symbol_L}
p (x, \xi) = -\sum\limits_{m=1}^{n} a_{2m} \xi^{2m} + q(x).
\end{equation}

It's obviously that $ p(x, \xi) $ belongs to the set $ S^{\, 2n} $
due to inequality
$$
\left| \frac{\partial^{k+l}}{\partial\xi^k \partial x^l} p(x, \xi)
\right| \le C_{k l} (1 +|\xi|)^{2n-k}, \quad k, l \in {\mathbf N}
\cup\{0\}.
$$
Moreover,
$$
\frac{\partial^{\,l}}{\partial x^l} \, p(x, \xi) \in S_0^{2n}, \quad
l \in{\mathbf N}.
$$

According to assumptions of theorem 1 the operator 
$ L: \mathrm{H}_{s+2n} ({\mathbf R}) \rightarrow \mathrm{H}_{s}
({\mathbf R}) $ satisfies all conditions of theorem 2 for any $ s
\in {\mathbf R} $. So, it is the Noether operator. As consequence
the operator $ L: \mathrm{H}_{s + 2n} ({\mathbf R}) \to
\mathrm{H}_{s} ({\mathbf R}) $ is normally solvable.

By $ L^* $ denote an operator being adjoint to the operator $ L $.

Let us assume kernel of the operator $ L^* $ be nontrivial. Then
differential equation (\ref{high_eq}) has a solution in the space $
\mathrm{H}_{s} ({\mathbf R}) $ if and only if the following
condition of orthogonality
\begin{equation} \label{adjoint}
<f, ker L^*> = 0
\end{equation}
holds.

Since
$$
L^* = - \sum\limits_{m=1}^n a_{2m} \frac{d^{2m}}{d x^{2m}} + q(x),
$$
then $ ker (L^*) \subset \bigcap\limits_{s \in {\mathbf R}}
\mathrm{H}_s({\mathbf R})$ \cite{GR1}.

Using Sobolev embedding theorems for the spaces $ \mathrm{H}_{s}
({\mathbf R}) $, $ s \in {\mathbf R} $, we find $ v_0^* \in \bar
{\mathrm{C}}_0^{\infty}({\mathbf R}) $ for any element $ v_0^* \in
ker {L^*} $.

As a consequence of the orthogonality condition (\ref{adjoint}) and
theorem 2 one obtains that the solution $ v(x) $ of equation
(\ref{high_eq}) belongs to space $ \bigcap\limits_{s \in {\mathbf
R}} \mathrm{H}_s({\mathbf R})$.

Arguing as above we get $ v \in \bar {\mathrm{C}}_0^{\infty}
({\mathbf R}). $

Now let us show that moreover $ v \in \mathrm{S}({\mathbf R}) $.
Indeed, since the function $ v \in \bar {\mathrm{C}}_0^{\infty}
({\mathbf R}) $ and it satisfies equation
\begin{equation}\label{quikly_decres_fun}
- \sum\limits_{m=1}^n a_{2m} \frac{d^{2m} \, v}{d x^{2m}} = - q v +
f,
\end{equation}
where the function $ - q v + f \in \mathrm{S}({\mathbf R}) $, then
due to properties of elliptic pseudodifferential operators with
polynomial coefficients \cite{GR2} we deduce that any solution to
equation (\ref{quikly_decres_fun}) from the space $
\mathrm{S}^*({\mathbf R}) $ belongs to the space $
\mathrm{S}({\mathbf R}) $. Thus, $ v \in \mathrm{S}({\mathbf R}) $.

Continuing this line of reasoning we see $ v_0^* \in
\mathrm{S}({\mathbf R}) $. Kind of the operator $ L^* $ implies
equality $ v_0^* = v_0 $. It means that orthogonality condition
(\ref{adjoint}) is equivalent to the one (\ref{ort_cond}).

From the above consideration it also follows that if the kernel of
the operator $ L : \mathrm{S}({\mathbf R}) \to \mathrm{S}({\mathbf
R}) $ is trivial, i.e. the homogeneous equation $ L v = 0 $ has the
only trivial solution in the space $ \mathrm{S}({\mathbf R}) $, then
equation (\ref{high_eq}) has a solution in the space $
\mathrm{S}({\mathbf R}) $ for any $ f \in \mathrm{S}({\mathbf R}) $.

The theorem 1 is proved.

\section{Conclusions}
Theorem on existence of a solution to the linear non-homogeneous
differential equation in the Schwartz space is proved. The theorem
can be used while studying soliton solutions to integrable systems
of modern mathe\-ma\-ti\-cal physics \cite{BPS} and asymptotic
soliton like solutions to singular perturbed higher order partial
differential equations of integrable type with variable
coefficients.

The theorem 1 generalizes the statement on existence of a solution
to the non-homogeneous equation with the one-dimensional Schrodinger
operator in the space of quickly decreasing functions
\cite{SamShrod}.

\section{Acknowledgements}
The authors are greatly thankful to Prof. Anatoliy Prykarpatsky for
the discussions and useful remarks.

This research was partially supported by Ministry of Education and
Science of Ukraine and Taras Shevchenko National University of Kyiv
[grant number 16~BA~038~--~01].

\end{document}